\documentclass[12pt]{article}
\usepackage{bbm}

 \def\be{\nopagebreak[3]\begin{equation}}
\def\ee{\end{equation}}
\def\ba{\nopagebreak[3]\begin{eqnarray}}
\def\ea{\end{eqnarray}}

\def\l{\langle}
\def\r{\rangle}
\newcommand{\y}{\hat{y}}              

\newcommand{\py}{{\hat{\pi}^{(y)}}}     
\newcommand{\pyRI}{\hat{\pi}^{(y)R,I}}

\newcommand{\pf}{\hat{\pi}}           
\newcommand{\fluc}[1]{(\Delta #1)^2}      
 
\newcommand{\cre}{\hat{a}^{\dagger}}    

\newcommand{\ann}{\hat{a}}            

\begin{document}
\title{A  signature of quantum gravity at the source of  the seeds of cosmic structure?}

\author{Daniel Sudarsky\\
Instituto de
Ciencias Nucleares\\
Universidad Nacional Aut\'onoma de M\'exico\\
A. Postal 70-543, M\'exico D.F. 04510, M\'exico}
\maketitle


\begin{abstract}
This  article reviews a recent work  by  a couple of colleagues and myself \cite{InflationUS} 
  about the shortcomings of the standard explanations  of the quantum   origin of cosmic  structure  in the inflationary scenario, and a proposal to address them. The  point it that in the usual accounts the inhomogeneity and anisotropy of  our universe  seem to emerge from an exactly homogeneous and isotropic initial state through processes that do not break those symmetries.   
  We  argued that some novel  aspect of  physics must be called upon to able to address the problem
  in a fully satisfactory way.  The  proposed approach  is inspired on Penrose's ideas regarding  an quantum gravity induced,  real and dynamical collapse of the wave function.    
\end{abstract}


\section{Introduction}
\medskip

One of the most important advances in physical cosmology  are the precision  measurements of the anisotropies in the CMB together with 
 their explanation within the context of the inflationary scenarios.
However after  the first glances at the  explanations one notices something odd:
 The description of our Universe starts with an initial set of  conditions which are
  totally homogeneous and isotropic both in the background space-time and in the quantum state that is  supposed to describe the "fluctuations", and it is quite easy to 
 see that the subsequent evolution through dynamics that do not break these symmetries can only lead 
 to an equally homogeneous and anisotropic universe. The arguments normally used in order to deal with  this issue, are  phrased in terms of ``the quantum to classical transition",  without focussing on the required 
breakdown of homogeneity and isotropy in the state.  
Thinking   in terms of first principles, 
one would start by acknowledging that the correct description of the  problem at hand would involve a full theory of quantum gravity coupled to a theory of  all the matter quantum  fields, and that there, the issue would be  whether we start with a quantum state that is homogeneous and isotropic or not?. If one chose to ignore  the problem and view it as something inherent to our approximations, one  could not argue that  one has an  understanding the origin of the CMB spectrum. 

   Penrose's \cite{Penrose}, ideas regarding the fundamental
 changes,  that he argues are needed in  quantum mechanics and their connection to quantum gravity, are used as inspiration 
  in the  treatment developed in \cite{InflationUS}:  The idea is  brings up the aspect  that we view as  part of the
 quantum gravity realm, to the forefront in order to modify-- in a minimalistic way-- the semiclassical treatment, to deal with the unsatisfactory part of the standard inflationary accounts of the issues. 

 \section{ The quantum origin of the seeds of cosmic structure}\label{sec_main}
\medskip
  Most  colleagues who have been working in this 
field f take the view,  that there is no problem at all  in the transition from a homogeneous and isotropic early  state of the universe,  a late state that is neither.  It is however a fact that the arguments  invoked in this regard  tend to differ  from one inflationary cosmologist
 to another \cite{Cosmologists}.  Very few do acknowledge that there seems to be something unclear at this point \cite{Cosmologists2}.
  One can see that the  situation  at hand, is quite different from any other situation usually treated using quantum mechanics where  the theory affords, at least one  
self consistent assignment 
at  each time of a state of the Hilbert space to our physical system, at each time. In trying to the consider
  such
 assignment when presented  with any of the proposed  justifications offered to deal with the issue 
one must be ready  to accept one of the
 following: i)
our universe was not really 
in that symmetric state (corresponding to the vacuum of the quantum field), ii) our universe is still described by a symmetric state, 
iii) at least at some points in the past the description of the state of our universe could not be done within quantum mechanics, iv) 
quantum mechanics does not correspond to
the full description of  a system at all times,  or v) our own observations of the
 universe mark the transition from a symmetric to an asymmetric state. None of these  represent 
a satisfactory alternative,
in particular, if we want to claim, that we understand the  
evolution of our universe and its structure -- including ourselves -- as the result of the fluctuations of quantum origin in 
the very early stages of 
our cosmology. 


Next  we  give a short description of this analysis. The
staring point is as usual the action of a scalar field coupled to
gravity. 
\be
\label{eq_action}
S=\int d^4x \sqrt{-g} \lbrack {1\over {16\pi G}} R[g] - 1/2\nabla_a\phi
\nabla_b\phi g^{ab} - V(\phi)\rbrack,
\ee
 where $\phi$ stands for the inflaton and for   its  potential $V$.  One then splits both, metric and
scalar field into a spatially homogeneous "background'" part and an
inhomogeneous part "`fluctuation'", i.e. the scalar  field is written
$\phi=\phi_0+\delta\phi$, while the  perturbed metric can, (after appropriate gauge fixing and by focussing on the scalar perturbation) be written
\begin{equation}
ds^2=a(\eta)^2\left[-(1+ 2 \Psi) d\eta^2 + (1- 2
\Psi)\delta_{ij} dx^idx^j\right],
\end{equation}
 where $\Psi$  is the relevant perturbation  called
the "Newtonian potential".

 The background solution
 corresponds to the standard inflationary cosmology  
 during the inflationary era  has a scale factor
$a(\eta)=-\frac{1}{H_{\rm I} \eta},$
 with $ H_I ^2\approx  (8\pi/3) G V$while the scalar $\phi_0$ field in the slow roll regime so $\dot\phi_0= - ( a^3/3 \dot a)V'$.

The perturbation of the scalar field leads to a perturbation of the energy momentum tensor, and
thus Einstein's equations at lowest order lead to
\begin{equation}
\nabla^2 \Psi  = 4\pi G \dot \phi_0 \delta\dot\phi  \equiv s \delta\dot\phi
\label{main3}
\end{equation}
where $s=4\pi G \dot \phi_0$.  This will be our main equation.  Next,  we write the quantum theory of the rescaled  the  field  $y=a \delta \phi$. We
consider the field in a box of side $L$, and  write

We rewrite the field and momentum operators  as
\begin{equation}
\y(\eta,\vec{x})=
 \frac{1}{L^{3}}\sum_{\vec k}\ e^{i\vec{k}\cdot\vec{x}} \hat y_k
(\eta), \qquad \py(\eta,\vec{x}) =
\frac{1}{L^{3}}\sum_{\vec k}\ e^{i\vec{k}\cdot\vec{x}} \hat \pi_k
(\eta),
\end{equation}
where the sum is over the wave vectors $\vec k$ satisfying $k_i L=
2\pi n_i$ for $i=1,2,3$ with $n_i$ integers, and  
where $\hat y_k (\eta) \equiv y_k(\eta) \ann_k +\bar y_k(\eta)
\cre_{-k}$ and  $\hat \pi_k (\eta) \equiv g_k(\eta) \ann_k + \bar g_{k}(\eta)
\cre_{-k}$
with
\begin{equation}
y^{(\pm)}_k(\eta)=\frac{1}{\sqrt{2k}}\left(1\pm\frac{i}{\eta
k}\right)\exp(\pm i k\eta),\qquad
g^{\pm}_k(\eta)=\pm
i\sqrt{\frac{k}{2}}\exp(\pm i k\eta) . \label{Sol-g} 
\end{equation}
Given that  we are  interested in considering a kind of self induced collapse which
 operates in close analogy with  a ``measurement",  we write the decompositions $\hat y_k (\eta)=\hat y_k{}^R
(\eta) +i \hat y_k{}^I (\eta)$ and $\hat \pi_k (\eta) =\hat \pi_k{}^R
(\eta) +i \hat \pi_k{}^I (\eta)$ where the operators $\hat y_k^{R, I} (\eta)$ and $\hat
\pi_k^{R, I} (\eta)$ are  hermitian.  
Next we provide a simple   specification of  what we mean by  ``the  collapse of the wave function"  by stating the form collapsed state  in terms of its collapse time.   We assume   the collapse to 
be  analogous to  some sort of  imprecise measurement of the
operators $\hat y_k^{R, I}
(\eta)$ and $\hat \pi_k^{R, I} (\eta)$.  
 Let $|\Xi\rangle$ be any state in the Fock space of
$\hat{y}$,and assign to each such state  the following quantity:
$d_k^{R,I}= \l \ann_k^{R,I} \r_\Xi.$
The expectation values of the modes of the fundamental field operators are then  expressible
as
\begin{equation}
\l {\y_k{}^{R,I}} \r_\Xi = \sqrt{2} \Re (y_k d_k^{R,I}),  \qquad
\l {\py_k{}^{R,I}} \r_\Xi = \sqrt{2} \Re (g_k d_k^{R,I}).
\end{equation}

For the vacuum state $|0\rangle$ we  have of course:
$
\l{\y_k{}^{R,I}}\r_0 = 0, \l\py_k{}^{R,I}\r_0 =0,
$
while their corresponding uncertainties are
\begin{equation}\label{momentito}
\fluc{\y_k {}^{R,I}}_0 =(1/2) |{y_k}|^2(\hbar L^3), \qquad
\fluc{\pf_k {}^{R,I}}_0 =(1/2)|{g_k}|^2(\hbar L^3).
\end{equation}
{\bf The collapse:}
In order to describe is the state $|\Theta\rangle$ after the
collapse we must  specify 
$d^{R,I}_{k} = \langle\Theta|\ann_k^{R,I}|\Theta\rangle $.

This is done by making the following
assumption about the state $|\Theta\rangle$ after
collapse:
\begin{eqnarray}
\l {\y_k^{R,I}(\eta^c_k)} \r_\Theta&=&x^{R,I}_{k,1}
\sqrt{\fluc{\y^{R,I}_k}_0}=x^{R,I}_{k,1}|y_k(\eta^c_k)|\sqrt{\hbar L^3/2},\\
\l {\py_k{}^{R,I}(\eta^c_k)}\r_\Theta&=&x^{R,I}_{k,2}\sqrt{\fluc{\pyRI_k}
_0}=x^{R,I}_{k,2}|g_k(\eta^c_k)|\sqrt{\hbar L^3/2},
\end{eqnarray}
where $x_{k,1},x_{k,2}$ are  selected randomly from within a Gaussian
distribution centered at zero with spread one.
From these equations we  solve for $d^{R,I}_k$.
We note that our universe, corresponds 
to a single realization of the random variables, and thus  each of the quantities 
$ x^{R,I}{}_{k,1,2}$ has a  single specific value. 
Later, we will see how to make relatively specific predictions, despite  these features.

The gravitational sector is treated at the semiclassical level so basic formula
Eq.(\ref{main3}) turns into
\begin{equation}
\nabla^2 \Psi = s \langle \delta\dot\phi\rangle. \label{main4}
\end{equation}
Before the collapse occurs, the expectation value on the right hand
side is zero. Let us now determine what happens after the collapse: To
this end, take the Fourier transform of  Eq.(\ref{main4}) and obtain
\begin{equation}\label{modito}
\Psi_k(\eta)=\frac{-s}{k^2}\langle \delta\dot\phi_k\rangle_\Theta=\frac{-s}{k^2}\sqrt{\hbar L^3 k}\frac{1}{2a}F(k), \label{F}.
\label{Psi}
\end{equation}
where
\begin{equation}
F(k) = (1/2) [A_k (x^{R}_{k,1} +ix^{I}_{k,1}) + B_k (x^{R}_{k,2}
+ix^{I}_{k,2})],
\end{equation}
with
\begin{equation}  A_k =  \frac {\sqrt{ 1+z_k^2}} {z_k} \sin(\Delta_k) , \qquad  B_k
=\cos (\Delta_k) + (1/z_k) \sin(\Delta_k),
\end{equation}
and where  $\Delta_k= k \eta -z_k$ with $ z_k =\eta_k^c
k$.

  Next we turn to the  observational quantities. We will,  disregard the changes to
dynamics that happen after re-heating  and due to the transition to
standard (radiation dominated) evolution. The quantity that is measured is ${\Delta T \over T}
(\theta,\varphi)$ which is a function of the coordinates on the
celestial two-sphere which is expressed as $\sum_{lm} \alpha_{lm}
Y_{l,m}(\theta,\varphi)$.  The angular variations of the 
temperature are then identified with the corresponding variations in the
``Newtonian Potential" $ \Psi$, by the understanding that they are the
result of gravitational red-shift in the CMB photon frequency $\nu$ so
${{\delta T}\over T}={{\delta \nu}\over {\nu}} = {{\delta (
    \sqrt{g_{00}})}\over {\sqrt{g_{00}}}} \approx \Psi$.  Thus , the
  measured quantity is the
``Newtonian potential" on the surface of last scattering: $
\Psi(\eta_D,\vec{x}_D)$,  from where one 
extracts
$
a_{lm}=\int \Psi(\eta_D,\vec{x}_D) Y_{lm}^* d^2\Omega.
$
To evaluate the  expected value for the quantity of interest we use (\ref{Psi}) and (\ref{F}) to
write
\begin{equation}
 \Psi(\eta,\vec{x})=\sum_{\vec k}\frac{s  } {k^2}\sqrt{\frac{\hbar
k}{L^3}}\frac{1}{2a}
 F(\vec{k})e^{i\vec{k}\cdot\vec{x}},
\label{Psi2}
\end{equation}

then, after some algebra we obtain
\begin{eqnarray}
\alpha_{lm}&=&s\sqrt{\frac{\hbar}{L^3}}\frac{1}{2a} \sum_{\vec
k}\frac{U(k)\sqrt{k}}{k^2} F(\vec k)  4 \pi i^l  j_l((|\vec k|
R_D) Y_{lm}(\hat k),\label{alm1}
\end{eqnarray}
where $\hat k$ indicates the direction of the vector $\vec  k$. It is in this
expression that the justification for the use of statistics becomes
clear.  The quantity we are in fact considering is the result of 
the combined contributions of an
ensemble of harmonic oscillators each one contributing with a complex
number to the sum, leading to what is in effect a 2 dimensional random
walk whose total displacement corresponds to the observational
quantity. Next we evaluate the most likely value
for such total displacement with the help of the imaginary
ensemble of universes, and the identification of the most likely value
with the  mean ensemble vale. After taking the continuum limit 
and  rescaling ing the variables of integration to $x =kR_D$, we find 
\begin{equation}
|\alpha_{lm}|^2_{M. L.}=\frac{s^2   \hbar}{2 \pi a^2} \int
\frac{C(x/R_D)}{x^4}    j^2_l(x) x^3 dx,
\label{alm5}
\end{equation}
where  
\begin{equation}
C(k)\equiv 1+ (2/ z_k^2) \sin (\Delta_k)^2 + (1/z_k)\sin (2\Delta_k).
\label{ExpCk}
\end{equation}
In the exponential expansion regime where $\mu$ vanishes and in
the limit $z_k\to -\infty$ where $C=1$, we find:
 \begin{equation}
 |\alpha_{lm}|^2_{M. L.}=\frac{s^2    \hbar} {2  a^2}
\frac{1}{l(l+1)} .
\end{equation}
 which has the standard  functional result.  However we must consider  the effect of the finite value of times of
collapse $\eta^c_k$ codified in  the function
$C(k)$. We note is that in order to get a reasonable
spectrum there  is a single  simple option: That $z_k $ be
essentially independent of $k$ that is the time of collapse of the
different modes should depend on the mode's frequency according to
$\eta_k^c=z/k$. This is a rather strong conclusion which could  represent   relevant information about whatever the mechanism of collapse is. 
 
\section{ A version of  `Penrose's  mechanism' for collapse in the cosmological  setting}
\label{sec_penrose}
\medskip
Penrose has argued   that  the collapse of quantum
mechanical wave functions is  dynamical process independent of observation, and that the
underlying mechanism is related to  quantum gravity. More precisely, according to this suggestion, the collapse
into one of several coexisting  quantum
mechanical alternatives would take place when the gravitational
interaction energy between the alternatives exceeds a certain
threshold.
A naive realization of Penrose's ideas in the present setting
could be obtained as follows: Each mode would collapse by the
action of the gravitational interaction between it's own possible
realizations. In our case, one could estimate the interaction energy
$E_I(k,\eta)$ by considering two representatives of the possible
collapsed states on opposite sides of the Gaussian associated with
the vacuum. We interpret $\Psi$, literally as the Newtonian
potential and consequently , $\rho= a^{-2}\dot\phi_0 \delta\dot\phi $ should  be identified  with  matter density. Then for the interaction energy between alternatives   we  would have:
\be\label{GE1}
E_I(\eta)=\int \Psi^{(1)} \rho^{(2)}dV = (a/L^3)\dot\phi_0 \Sigma_{k}
\Psi^{(1)}_{ k}( \eta) \delta\dot\phi^{(2)}_{k}(\eta) , 
\ee 
where $(1),(2)$
refer to the two different realizations chosen. Recalling
 that $\Psi_{ k} = ( -s/k^2)  \delta\dot\phi_k$,   we find 
  \be
 E_I(\eta)= -4\pi G (a/L^3)
\dot\phi_0^2\Sigma_{k} (1/k^2)
 \delta\dot\phi^{(1)}_{k}(\eta) \delta\dot\phi^{(2)}_{k}(\eta)\approx \Sigma_{k}( \pi \hbar  G/ak) (\dot\phi_0)^2.
\ee 
  Where we have used equation (\ref{momentito}), to  estimate $ \delta\dot\phi^{(1)}_{k}(\eta) \delta\dot\phi_{k}^{(2)}(\eta) $ by 
 $|< \delta\dot\phi_k > |^2 = \hbar k L^3 (1/2a)^2$.

 This result can be interpreted as the  sum of  the contributions of each mode to the interaction energy of different alternatives.
  We  view each mode's collapse as occurring independently, so the collapse of mode  $k $ would  occur when this energy 
   $E_I(k,\eta)=( \pi \hbar  G/ak) (\dot\phi_0)^2 =\frac{ \pi \hbar  G } { 9H_I^2} ( a/k) ( V')^2$ reaches the  value of the Planck Mass $M_p$.  Thus the condition determining the time of collapse  $\eta^c_k$ of the mode $k$ becomes, 
\be 
  z_k=\eta^c_k k =\frac{\pi }{9} (\hbar V')^2(H_I M_p)^{-3}=\frac{\epsilon} {8\sqrt {6\pi}}(\tilde V)^{1/2}\equiv  z^c,
 \ee
which is independent of $k$,  and thus, as we saw in the previous  section  leads to a roughly  scale invariant spectrum of fluctuations in
accordance with observations. 

\section{Discussion}
\medskip

     First we address a recent article \cite{Keifer2}, which is part of this volume in which colleagues reiterate, in response to \cite{InflationUS}, that the problem we have alluded to, does not exist.  It is illuminating to consider their claims and  views in this regard:    1) That the situation is analogous to the spontaneous symmetry breaking in field theories, 2)  that the environment selects as preferential "pointer"  basis the  the one that diagonalizes the field  operators rather than the momenta operators because of couplings of other fields to the field operators, and 3) that "the initial symmetric vacuum state  evolves into a symmetric  superposition of inhomogeneous states out of which one component is {\it selected}".  The  first point is rather complex to discuss and will be addressed elsewhere,  limiting ourselves here to point out,  in general, the dangers of the "arguments by analogy" so lets focus here on the last two.   Point  2) seems to ignore the fact that most of the known interactions are of the guage-theory type and couple both to the momentum and the  other "spatial"   field gradients rather than to the underived  fields themselves. However  the clearest problem lies in   point 3) where the authors do acknowledge that the unitary evolution leads at late times to a ``symmetric  superposition of inhomogeneous states"  which is, according to quantum mechanics nothing but a  fully symmetric state.  Then somehow, { \it one of the components of this state gets ``selected"}. 
     Is this a physical  process or mechanism?,  does this  occur at some  time?,  if not,  then what is this { \it get selected} supposed to represent? Is this to be regarded as
       just part of our subjective perceptional framework?  if so what part of the treatment is not?  Do or do not the states
       represent the physical condition of the system they describe?  If they do not,why would we view the initial state as indicating the the early universe was homogeneous and isotropic? Could we say that we understand the origin of the anisotropies and inhomogeneities if we didn't claim  we started with a condition that was homogeneous and isotropic?. Perhaps  we should think  that our own actions play an active role in producing this selection?. If this is the case, are  we understand the emergence of the conditions that make us possible (the inhomogeneity and anisotropy of the universe) are in part  the result of our own actions? See also the issue of the assignment of a state at every time raised in the introduction.  In short, it is quite clear that something  strange
is being called upon with the statement that "one of the alternatives gets selected" which  character is not  being revealed  by avoiding to address  all these issues.  We must understand under which conditions that this selection mechanism operate. Our point of view is that it is always healthier to confront the hard issues face on, because even if one fails to find a satisfactory answer  at the start,  their 
acknowledgment is the first step to their eventual resolution. This is the  posture we have taken and we find it quite remarkable that
in doing so we are able to obtain a relatively satisfactory picture.  We do not know what exactly is the physics  of collapse but we
 were nevertheless able to obtain some constraints on it (about the time of collapse of the different modes), and shown that a simplistic extrapolation of Penrose«s ideas satisfy this constraint.

  In conclusion, we have  reviewed a serious  shortcoming of the inflationary account of the origin of cosmic structure, and have given a brief account of the proposals to deal with them which were 
 first reported in \cite{InflationUS}.
These  lines of inquiry have lead to the  recognition  that something else seems to be needed 
for the whole picture to work and that it  could  be pointing   towards an actual manifestation quantum gravity. 
We have shown that not only 
the issues are susceptible of scientific investigation based on observations, but also that a simple
 account of what is needed, seems to be
 provided by the extrapolation of Penrose's ideas to the cosmological setting. 
 Interestingly the scheme does in fact  lead to  some deviations from the standard picture where the metric and scalar field perturbations are quantized.  For one, as explained elsewhere\cite{Napflio},  one is lead to expect no excitation of the tensor modes
 because it is only the scalar  metric perturbation that  gets excited by the collapse of the quantum inflaton field.
 We also  find  new avenues to address the fine tuning problem that  affects most  inflationary models, because one can follow in more detail the  objects that give rise to the anisotropies and inhomogeneities, and by having  the possibility to consider independently the issues relative to formation of the perturbation, and their evolution through the
reheating era. That is, the present analysis offers a path to
 get rid of the ``fine tuning problem" for the inflationary scenarios
 \cite{InflationUS, Napflio}. Some of these aspects  can, in principle,  be tested, indicating that what initially could have been thought to be   essentially a philosophical problem, leads instead  to truly  physical issues.
   
Our main point is however that   in our  search for physical manifestations of new physics tied to quantum aspects of gravitation, we might be ignoring  the most dramatic such  occurrence:  The cosmic structure of the Universe itself.

\section*{Acknowledgments}

 \noindent
 It is a pleasure to acknowledge very  helpful conversations with J. Garriga, E. Verdaguer and A. Perez. This work was supported
 in part  by DGAPA-UNAM
IN108103 and CONACyT 43914-F grants.
 


\begin{thebibliography}{9}

\bibitem{InflationUS}
``On the Quantum Origin of the Seeds of Cosmic Structure", 
 A Perez, H Sahlmann and
 D Sudarsky,   {\it Class and Quant Gravity}  {\bf 23}  2317-54 (2006).
 

\bibitem{Penrose} 
``The Emperor's New Mind",
 R Penrose, {\it The Emperor's New Mind}, (Oxford
  University Press 1989).
 R. Penrose, On Gravity's  Role in Quantum State Reduction,
  in {\it  Physics meets philosophy at the Planck scale} Callender, C. (ed.) (2001).

\bibitem{Cosmologists}
``Decoherence in Quantum Cosmology", 
J J Halliwell,
{\it Phys. Rev. } D  {\bf 39}, 2912,(1989);
``Origin of Classical Structure From Inflation",
 C Kiefer
{\it Nucl.\ Phys.\ Proc.\ Suppl.} {\bf 88}, 255 (2000);
``Semiclasicallity and decoherence of Cosmological perturbations", 
D Polarski and A. A.
  Starobinsky [arXiv:gr-qc/9504030] (1996);
`` Environment Induced Superselection In Cosmology",
 W H Zurek, Environment Induced Superselection In Cosmology in {\it Moscow 1990, Proceedings, Quantum gravity} (QC178:S4:1990),  456-72 (see High Energy Physics Index 30 (1992) No. 624);
``Decoherence Funtional and Inhomogeneities in the Early Universe",
 R Laflamme and A Matacz {\it Int.\ J.\ Mod.\ Phys.  }D {\bf 2}, 171 (1993);
 ``The self-induced approach to decoherence in cosmology,''
  M Castagnino and O Lombardi, {\it  Int. J. Theor. Phys.} {\bf 42}, 1281, (2003);
``Decoherence during inflation: The generation of classical inhomogeneities,''
  F C Lombardo and D Lopez Nacir,
 {\it Phys.\ Rev. }D {\bf 72}, 063506 (2005);
``Inflationary Cosmological Perturbations of Quantum Mechanical Origin"
J Martin, {\it Lect.\ Notes Phys. } {\bf 669}, 199 (2005).

\bibitem{Cosmologists2}
T Padmanabhan, {\it Cosmology and Astrophysics Through Problems}, 
 (Cambridge University Press 1996);
J B Hartle,``Generalized Quantum mechanics for Quantum Gravity",  arXive: gr-qc/0510126.
 
\bibitem{Keifer2}" Pointer States for Primordial Fluctuations in Inflationary Cosmology",
C Keifer, I Lohmar,  D Polarski and A. S. Starobinsky,  arXive: astro-ph/0610700.

\bibitem{Napflio}"The Seeds of Cosmic structure as a door to New Physics",  D Sudarsky, arXive; gr-qc/0612005

\end{thebibliography}
\end{document}